\documentclass[11pt,preprint]{emulateapj}
\usepackage{psboxit}

\newcommand{\kms}{\,km~s$^{-1}$} 

\newcommand{\Msun}{\mbox{\,$M_{\odot}$}}

\def\spose#1{\hbox to 0pt{#1\hss}}
\def\simlt{\mathrel{\spose{\lower 3pt\hbox{$\mathchar"218$}}
     \raise 2.0pt\hbox{$\mathchar"13C$}}}
\def\simgt{\mathrel{\spose{\lower 3pt\hbox{$\mathchar"218$}}
     \raise 2.0pt\hbox{$\mathchar"13E$}}}

\PScommands

\newcommand{\Sersic}{S\'ersic}

\def\simless{\mathbin{\lower 3pt\hbox
	{$\,\rlap{\raise 5pt\hbox{$\char'074$}}\mathchar"7218\,$}}} 
\def\simgreat{\mathbin{\lower 3pt\hbox
	{$\,\rlap{\raise 5pt\hbox{$\char'076$}}\mathchar"7218\,$}}} 

\newcommand{\ndwarf}{9399~}  
\newcommand{\nfield}{2951~}    

\newcounter{thefigs}

\newcounter{thetabs}

\begin{document}

\title{A Stellar Mass Threshold for Quenching of Field Galaxies}

\author{M.\ Geha\altaffilmark{1},
M.~R.~Blanton\altaffilmark{2},
R. Yan\altaffilmark{2},
J.~L.~Tinker\altaffilmark{2}}

\altaffiltext{1}{Astronomy Department, Yale University, New Haven,
  CT~06520.  marla.geha@yale.edu}

\altaffiltext{2}{ Center for Cosmology and Particle Physics,
  Department of Physics, New York University, 4 Washington Place, New
  York, NY 10003}

\begin{abstract}
  We demonstrate that dwarf galaxies ($10^{7} < M_{\rm stellar} <
  10^9$\Msun, $-12 > M_r > -18$) with no active star formation are
  extremely rare ($<0.06$\%) in the field.  Our sample is based on the
  NASA-Sloan Atlas which is a re-analysis of the Sloan Digital Sky
  Survey Data Release 8.  We examine the relative number of quenched
  versus star forming dwarf galaxies, defining quenched galaxies as
  having no H$\alpha$ emission (EW$_{H\alpha} < 2 \mbox{\AA}$) and a
  strong 4000\mbox{\AA}-break.  The fraction of quenched dwarf
  galaxies decreases rapidly with increasing distance from a massive
  host, leveling off for distances beyond 1.5\,Mpc.  We define
  galaxies beyond 1.5\,Mpc of a massive host galaxy to be in the
  field.  We demonstrate that there is a stellar mass threshold of
  $M_{\rm stellar} < 1.0\times10^9$\Msun\ below which quenched
  galaxies do not exist in the field.  Below this threshold, we find
  that none of the \nfield field dwarf galaxies are quenched; all
  field dwarf galaxies show evidence for recent star formation.
  Correcting for volume effects, this corresponds to a 1-sigma upper
  limit on the quenched fraction of 0.06\%.  In more dense
  environments, quenched galaxies account for 23\% of the dwarf
  population over the same stellar mass range.  The majority of
  quenched dwarf galaxies (often classified as dwarf elliptical
  galaxies) are within 2 virial radii of a massive galaxy, and only a
  few percent of quenched dwarf galaxies exist beyond 4 virial radii.
  Thus, for galaxies with stellar mass less than $1.0\times10^9$\Msun,
  ending star-formation requires the presence of a more massive
  neighbor, providing a stringent constraint on models of star
  formation feedback.

\end{abstract}


\section{Introduction} \label{sec_intro}

A well-established color bimodality is seen in the local distribution
of luminous galaxies \citep[e.g.,][]{baldry06a, tanaka05a, blanton05b,
  blanton09a} and appears already in place at redshifts above $z\sim1$
\citep{bell04a,cooper07a}.  The galaxy population divides between
blue, star-forming systems and red, quenched systems.  The relative
fractions between these two populations depends on both stellar mass
and environment.  Luminous red galaxies exist both in the field and
denser regions, however, at fixed stellar mass the fraction of
luminous red galaxies is higher in denser environments
\citep[e.g.,][]{kauffmann04a,vandenbosch08a}.

The red/quenched fractions for less massive galaxies are lower
regardless of environment as compared to their higher mass
counterparts.  \citet{kauffman03b} found a characteristic stellar mass
of $3\times 10^{10}$\Msun\ below which galaxies tend to be star
forming and have lower surface brightnesses.  In the framework of
``central'' and ``satellite'' galaxies \citep{yang07a,Zehavi11a},
where central galaxies are defined as the most massive galaxy in their
dark matter halo based on group catalogs, \citet{wang09a} and
\citet{peng10a} showed that the red fractions for central galaxies
decrease as a smooth function of declining stellar mass, reaching red
fractions around 10\% for the lowest stellar masses ($M_{\rm stellar}
\sim10^9$\Msun) available in large numbers in the Sloan Digital Sky
Survey (SDSS).  Using spectroscopic diagnostics to define quenched
galaxies, \citet{wetzel10a} and \citet{tinker11a} also showed that
quenched fractions decrease with stellar mass for central galaxies,
reaching slightly lower quenched fractions for central galaxies at the
same stellar mass.

Many galaxy formation models predict lower quenched fractions for
lower mass central galaxies.  Currently favored models suggest that
massive galaxies are quenched because they can maintain a hot gaseous
halo, due to processes such as heating from supernova or active
galactic nuclei \citep[e.g.,][]{croton06a, dekel08a, kimm09a}.  These
processes suppress gas cooling, either heating gas within the galaxy
or preventing additional cold gas from accreting, and effectively
shutting off star formation \citep{keres05a}.   Low mass galaxies
which are satellites of more massive objects are quenched because of
the many physical mechanisms available in the group environment, such
as ram pressure, harassment and tidal stripping.  However, below some
threshold, small {\it central} galaxies are not expected to be
quenched \citep{gabor12a}.  Testing such a framework requires a large
well understood sample of low mass galaxies across a range of
environments.

Quantifying the quenched fractions of dwarf galaxies, defined here as
galaxies fainter than $M_r > -18$ or stellar mass below $10^9$\Msun,
is challenging, largely because the colors and apparent sizes of low
luminosity galaxies are similar to those of more luminous, and far
more numerous, background objects.  This necessitates spectroscopic
samples.  In the Local Group, quenched galaxies dominate the dwarf
satellite population within 500\,kpc of either the Milky Way or M31,
while star forming gas-rich galaxies tend to lie at larger distances
\citep{einasto74a,mateo98a, grcevich09a, weisz11a}.  Similarly, in the
nearby Virgo or Coma galaxy clusters, quenched dwarf galaxies dominate
the clusters' center and transition to a more star forming population
in the outskirts \citep{binggeli85b,ferguson91a, lisker07a}.
\citet{haines08a, haines07a} found no quenched galaxies fainter than $M_r =-18$
in the SDSS Data Release 4.  In this paper, we perform a similar
analysis, confirming this result with a much larger sample of dwarf
galaxies.

We construct a clean sample of nearly 10,000 dwarf galaxies in the
stellar mass range $10^7 < M_{\rm stellar} < 10^9$\Msun\ ($-12 > M_r >
-18$) across a wide range of environments, using a re-analysis of the
SDSS Data Release 8.    We demonstrate that quenched galaxies which have
ceased forming stars, i.e., ``red and dead'' galaxies, often
classified as dwarf elliptical or dwarf spheroidal galaxies, do not
exist in isolation (or as ``central'' galaxies) for stellar masses
below $1.0\times 10^{9}$\Msun.  In \S\,\ref{sec_data}, we discuss the
sample construction and environmental parameterization using a nearest
neighbor distance.  In \S\,\ref{ssec_environ}, we demonstrate the
absence of quenched dwarf galaxies passing our isolation criteria,
noting a significant fraction of isolated red star forming galaxies.
In \S\,\ref{ssec_thresh}, we compare the fraction of field quenched
galaxies as a function of stellar mass.  In \S\,\ref{ssec_virial}, we
compare the distribution of quenched and star forming dwarf galaxies
as a function of distance from its host galaxy, in units of the host’s
virial radius.  In \S\,\ref{ssec_SF}, we briefly examine the
properties of the star forming field dwarf galaxy sample.  We conclude in
\S\,\ref{sec_concl} with the implication of these results for star
formation feedback and galaxy formation.

Throughout this paper, we assume cosmological parameters $\Omega_0 =
0.3$, $\Omega_\Lambda = 0.7$, and $H_0 = 70$ km s$^{-1}$
Mpc$^{-1}$.  All magnitudes in this paper are $K$-corrected to
rest-frame bandpasses using the method of \citet{blanton06a} and {\tt
  kcorrect} {\tt v4\_2}.  Because of the small range of look-back
times in our sample (a maximum of around 700 Myr), we do not
evolution-correct any of our magnitudes.

\begin{figure}
\epsscale{1.2}
\plotone{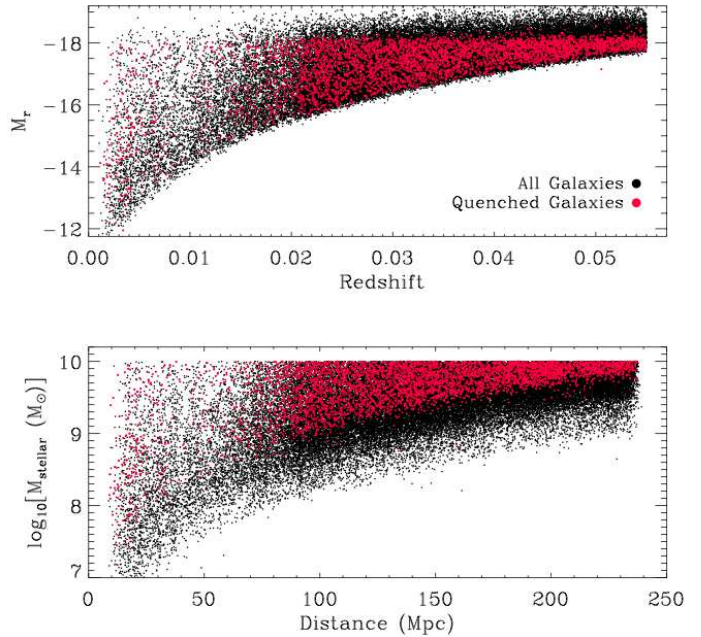}
\caption{The NASA-Sloan Atlas (NSA) galaxy sample showing ({\it
    top}) absolute magnitude versus redshift and ({\it bottom})
 stellar mass versus distance.  The NSA has an arbitrary
  redshift cutoff at $z = 0.055$.  For the purposes of this paper, we
  analyze only galaxies with stellar mass less than $10^{10}$\Msun.
  Quenched galaxies (red symbols) are intrinsically less luminous at a
  given stellar mass and are therefore found over a smaller volume as
  compared to star-forming galaxies at the same stellar mass.  We
  account for this difference using 1/V$_{\rm max}$
  corrections.  \label{fig_zdist}}
\end{figure}

\vskip 0.5cm 
\section{ Constructing the SDSS Dwarf Galaxy Sample}
\label{sec_data}

\subsection{The NASA-Sloan Atlas}

Our dwarf galaxy sample is derived from the SDSS Data Release 8
spectroscopic catalog \citep[DR8; ][]{dr8}, covering
$7966~\mathrm{deg^2}$ of the sky.  Since the SDSS main catalog is not
optimized for nearby low luminosity objects, we instead select objects
from the NASA-Sloan Atlas\footnote{\tt http://www.nsatlas.org}(NSA).
The NSA is a reprocessing of the SDSS photometry using the SDSS
$ugriz$ images with an improved background subtraction technique
(\citealt{blanton11a}), combined with GALEX images in the near and
far-UV.  The NSA photometry is a significant improvement over the
standard SDSS DR8 photometric catalog, as described in
\citet{blanton11a}.   All galaxies with redshifts $z<0.055$ within the
SDSS footprint are analyzed (Figure~\ref{fig_zdist}).  For each galaxy, the NSA contains a
mosaicked image which is deblended and analyzed consistently in all
bands.  Fluxes are based on two-dimensional \Sersic\ models whose
structural parameters are fit to the $r$-band image.  The NSA catalog
galaxy also provides a re-analysis of the SDSS spectroscopic data for
each galaxy using the techniques of \citet{yan11a} and the SDSS
spectrophotometric recalibration of \citet{yan11b}.  This analysis
yields fluxes, equivalent widths and associated errors.

To estimate stellar masses, we use those reported by the {\tt
  kcorrect} software of \citet{blanton06b} which assumes a
\citet{chabrier03a} initial mass function and are based on fits to
both the SDSS optical and, when available, GALEX fluxes.  Distances
are estimated based on the SDSS NSA redshift and a model of the local
velocity field \citep{willick97a}.  Distance errors are folded into
our error estimations for stellar mass.

In this paper, we focus largely on dwarf galaxies with stellar masses
between $10^{7} < M_{\rm stellar} <10^{9} M_\odot$.  In the NSA
catalog this corresponds to \ndwarf galaxies.  A search for objects in
the same stellar mass and redshift range in the DR7 NYU Value-Added
Galaxy Catalog \citep[VAGC; ][]{blanton05a} yields nearly 16,000
objects over the same area.  The standard SDSS photometry used by the
VAGC catalog is not optimized for extended nearby galaxies; the excess
of 'dwarfs' in this catalog are primarily shredded pieces of massive
galaxies which have been properly accounted for in the NSA catalog.

\begin{figure}[t]
\epsscale{1.2}
\plotone{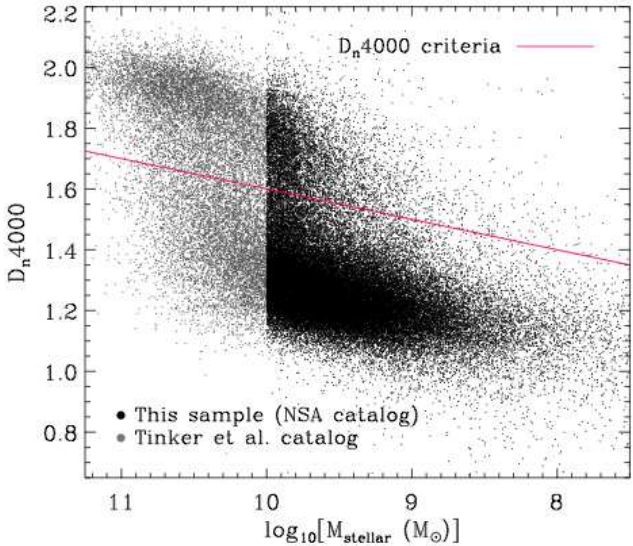}
\caption{D$_n$4000 versus stellar mass for our
  dwarf galaxy sample (black symbols) and the \citet{tinker11a} group
  catalog (grey symbols).  In the overlap region between $10^{9.6 -
    10}$\Msun, the two catalogs agree well.  We define quenched
  galaxies as having both EW H$\alpha < 2$\mbox{\AA} and
  D$_n$4000 greater than the red line in this
  plot.  \label{fig_d4000}}
\end{figure}

\begin{figure*}[t!]
\epsscale{1.15}
\plotone{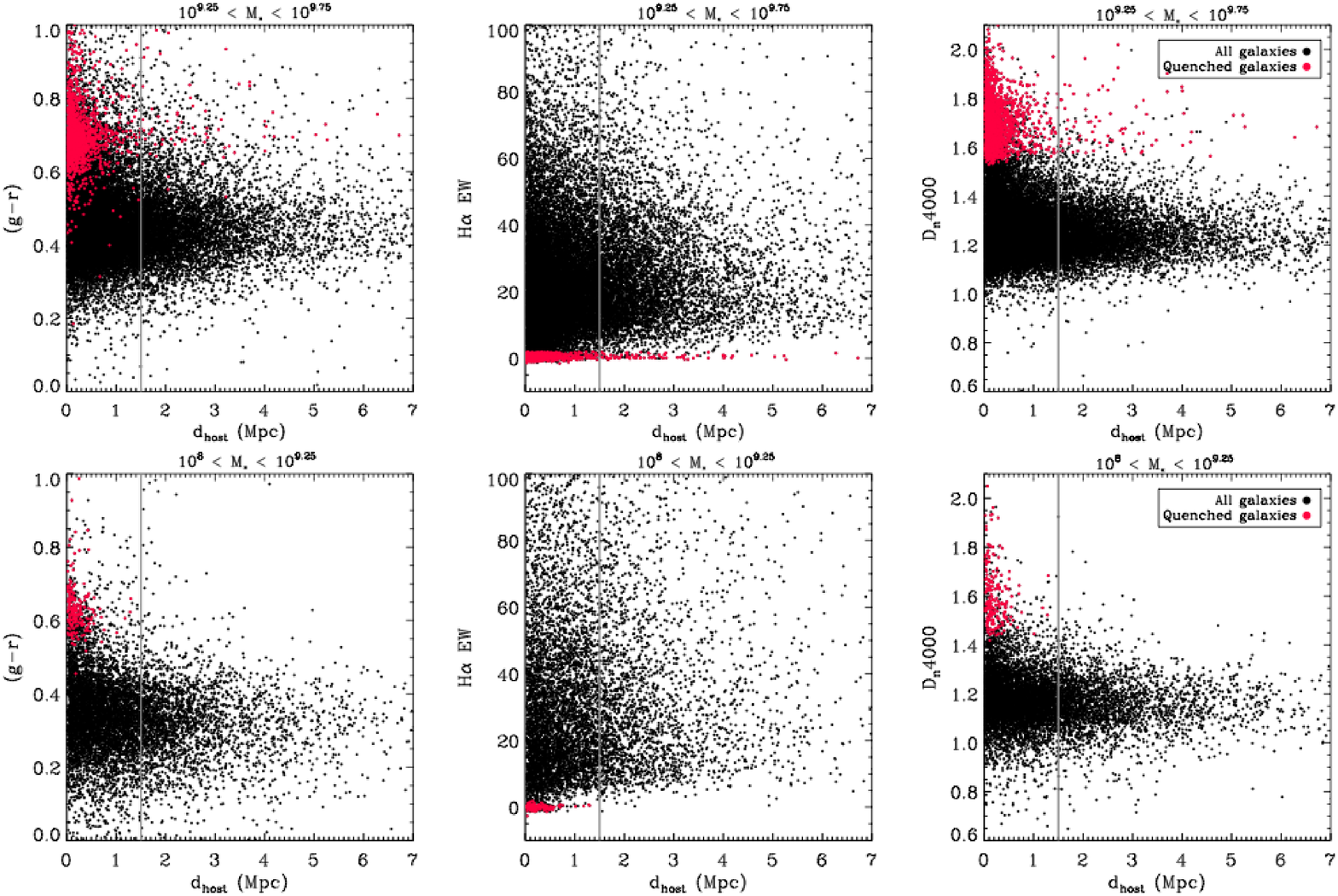}
\caption{\label{fig_3panels} The distribution of NSA galaxies in the
  mass range $10^{9.25} - 10^{9.75}$\Msun ({\it top}) and $10^8 -
  10^{9.25}$\Msun ({\it bottom}).  Galaxies are plotted as a function
  of nearest neighbor distance $d_{\rm host}$ versus $g-r$ color ({\it
    left}), H$\alpha$ equivalent width ({\it middle}), and D$_n$4000
  ({\it right}).  Red symbols indicate quenched galaxies passing our
  H$\alpha$ and D4000 criteria.  Galaxies with $d_{\rm host} >
  1.5$\,Mpc (solid gray line) are considered isolated.}
\end{figure*}

\subsection{Environment and Nearest Luminous Neighbors}

We calculate environments for our dwarf galaxy sample relative to more
luminous neighbor galaxies.  We want to maximize our ability to
identify luminous neighbors, despite the large angular distances for this nearby
sample.  For example, searching a 1\,Mpc region around a
galaxy 30\,Mpc away corresponds to 2 degrees on the sky.  Many of our
dwarf galaxies are on the SDSS Southern stripes, which are only 2.5
degrees wide.  For this reason, instead of the NSA catalog, which is
limited to the area with SDSS imaging, we use the 2MASS Extended
Source Catalog to identify luminous galaxy hosts.  For redshifts, we
use SDSS spectroscopy plus several other sources: the 2-degree Field
Galaxy Redshift Survey (2dFGRS; \citealt{colless01a}), the 6-degree
Field Galaxy Redshift Survey (6dfGRS; \citealt{jones04a}),
ZCAT\footnote{\tt
  http://www.cfa.harvard.edu/$\sim$dfabricant/huchra/zcat/}, ALFALFA
(\citealt{giovanelli05a}) as well as every redshift within $z<0.055$
from the NASA Extragalactic Database (NED). By using this all-sky
catalog, we can best identify luminous neighbors even for dwarf
galaxies near the edges of the SDSS imaging.

To quantify environment, we determine the distance, $d_{\rm host}$,
for each of our dwarf galaxies to its nearest ``luminous'' neighbor.
In this context, we define galaxies as luminous if $M_{K_s} <-23$
corresponding to a stellar mass of approximately 2.5$\times
10^{10}$\Msun\ (assuming a mean stellar mass-to-light ratio in the
$K_s$-band of unity).  The 2MASS sample is complete within our dwarf
galaxy volume for this choice of $M_{K_s}$.  Because we use several
different redshift surveys, the sample of luminous galaxies is
non-uniform, especially outside the area covered by SDSS main sample
spectroscopy. However, we prefer to have as complete a sample of
luminous galaxies as possible, in order to most reliably identify
isolated dwarf galaxies.  To define the environment for each dwarf
galaxy, we search for the closest luminous galaxy within 1000\kms\ in
redshift and within a projected comoving distance $d_{\rm
  host}<7$\,Mpc.  A small number of dwarfs have no luminous galaxy
within this volume; we set the $d_{\rm host}$ values for these
galaxies to 7\,Mpc.

\subsection{Definition of Quenched Dwarf Galaxies}

We divide our sample between galaxies with active star formation and
``quenched'' galaxies that are not forming stars.  We differentiate
between these two populations using SDSS spectroscopic diagnostics.
We define quenched galaxies as having both no H$\alpha$ emission (EW
H$\alpha< 2$\mbox{\AA}) and a criterion based on the 4000\mbox{\AA}
break, D$_n$4000.  The D$_n$4000 index is a measure of the
light-weighted age of the stellar population \citep{balogh99a}.
Because is it measured in two 100\mbox{\AA} windows separated by
50\mbox{\AA}, it is less affected by dust reddening than broad-band
galaxy colors.  Analogous to color indicators, we find that the
D$_n$4000 strength is a function of stellar mass
(Figure~\ref{fig_d4000}) and therefore define quenched galaxies as
having D$_n$4000$> 0.6 + 0.1$log$_{10}$(M$_{\rm stellar}[\Msun]$).  At
a stellar mass of $10^{10}$\Msun, this is equivalent to the
D$_n$4000$> 1.6$ criterion used by \citet{tinker11a}.

Figure~\ref{fig_3panels} shows $(g-r)$ color,
H$\alpha$ EW, and D$_n$4000 index as a function of nearest
neighbor distance, $d_{\rm host}$.  We compare the distributions of
these quantities for two stellar mass bins: $10^{9.25} < M_{\rm
  stellar} <10^{9.75} M_\odot$ (top panels), and $10^{8} < M_{\rm
  stellar} <10^{9.25} M_\odot$ (bottom panels).  These two stellar
mass bins were chosen to have a roughly similar number of galaxies and
straddle the stellar mass threshold of $1.0\times10^9$\,\Msun\
discussed in \S\,\ref{ssec_thresh}.  In both bins, the majority of
quenched galaxies have red ($g-r$) colors and tend to lie close to a
massive parent galaxy, at values of $d_{\rm host} < 1$\,Mpc.
Similarly, the majority of galaxies with strong D$_n$4000 index, an
indicator of older stellar populations, exist in close proximity to a
larger galaxy.  At all masses, galaxies show a wide range of H$\alpha$
EW which extend to values larger than the plotted region.  At higher
stellar masses, there are a number of galaxies which show no H$\alpha$
emission and high values of D$_n$4000 at large values of $d_{\rm
  host}$, however, these objects are missing in the lower mass panels.
We explore these quantities further below.

\subsection{V$_{\rm max}$ Corrections and Surface Brightness
  Completeness}\label{ssec_vmax}

The SDSS spectroscopic apparent magnitude limit of $r < 17.77$
restricts the volume over which dwarf galaxies can be found
(Figure~\ref{fig_zdist}).  We will compare quantities as a function of
stellar mass and do not want to use a volume-limited sample which
would further reduce the number of available dwarf galaxies.  Galaxies
in a given stellar mass bin will have a range of absolute magnitudes,
and therefore a range of volume over which they could be detected in
SDSS.  To account for this difference, we weight our sample using the
$1/V_{\rm max}$ method, determining the maximum volume $V_{\rm max}$
for which each galaxy could have been found, given the spectroscopic
apparent magnitude limit.  The NSA catalog includes galaxies with
redshifts less than $z < 0.055$, thus only galaxies with $M_r < -19.1$ are
found throughout the full sample volume.  We have compared our
$1/V_{\rm max}$ weighted results to that of a volume-limited sample
for objects with stellar mass greater than $10^8$\Msun.  The
volume-limited results are far noisier due to the significantly
smaller volume probed, but are qualitatively similar.

The SDSS is incomplete for low surface brightness galaxies.  The SDSS
spectroscopic survey completeness as a function of half-light surface
brightness drops below 50\% at $\mu_{50, r} \sim
23.5$\,mag~arcsec$^{-2}$ \citep{blanton04b}, and below 10\% at
$\mu_{50, r} = 24.0$\,mag~arcsec$^{-2}$.  For dwarf galaxies with
stellar mass below $10^9$\Msun, the median surface brightness of our
sample is $\mu_{50, r} = 22.3$\,mag~arcsec$^{-2}$.  Given the surface
brightness incompleteness, we would miss, for example, one out of the
five brightest satellites around the Milky Way \citep{mateo98a}.

For the purposes of this study, we are concerned only with whether or
not low surface brightness galaxies are preferentially missing from
our quenched sample, relative to the full catalog.  In dense
environments ($d_{\rm host} < 1$\,Mpc), where we detect both quenched
and star forming dwarf galaxies, the surface brightness distribution
of quenched systems peaks at slightly lower surface brightnesses
($\mu_{50, r} = 22.5$\,mag~arcsec$^{-2}$) as compared to the star
forming sample ($\mu_{50, r} = 22.3$\,mag~arcsec$^{-2}$).  However,
comparing the shape of the distributions via the Kolmogorov-Smirnov test
suggests that the quenched galaxies in dense environments could
plausibly be drawn from the surface brightness distribution of
star-forming galaxies ($P_{\rm KS} = 0.1$).  Unless there is a
population of quenched low surface brightness galaxies that exists only in the
field below the detection limits of SDSS, our red fractions should not
be biased by the surface brightness incompleteness of the survey.

\section{Results}

In the sections below, we calculate the quenched fraction, $f_{\rm
  quenched}$.  Weighting each galaxy by the total
volume over which it could be observed, the quenched fraction is:

\begin{equation}
f_{\rm quenched}  = \frac{\displaystyle\sum_{i=1}^{N_{\rm quenched}} 1/V_{\rm
    max}, i}
   {\displaystyle\sum_{i=1}^{N_{\rm quenched}  + N_{\rm SF}} 1/V_{\rm  max},i}
\end{equation}
where $N_{\rm quenched}$ and $N_{\rm SF}$ are the number of quenched
and star-forming galaxies, respectively.  We calculate 1-$\sigma$
errors for $f_{\rm quenched}$ by propagating Poisson counting
statistics errors on the independent quantities $N_{\rm quenched}$ and
$N_{\rm SF}$.  For cases where the number of quenched galaxies equals
zero, we calculate the 1-$\sigma$ upper limits directly according to
\citet{gehrels86a}.

\subsection{Quenched Fractions as a Function of Environment}\label{ssec_environ}

\begin{figure}[t!]
\epsscale{1.27}
\plotone{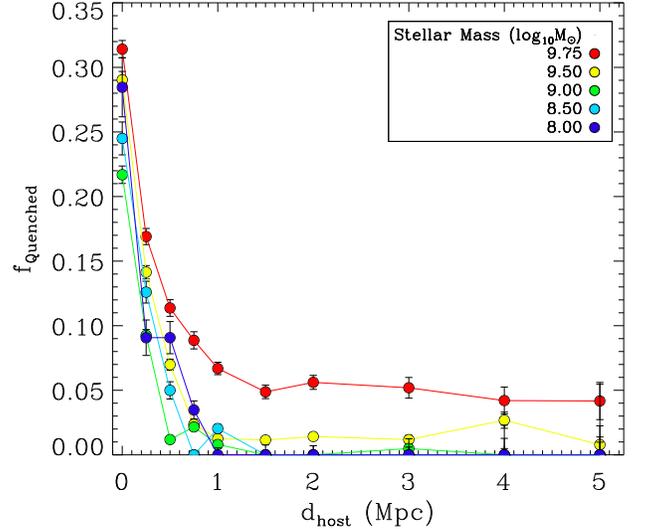}
\caption{\label{fig_environ} The fraction of quenched galaxies,
$f_{\rm quenched}$, as a function of distance to the nearest luminous
galaxy, $d_{\rm host}$, in several stellar mass bins.  We define
'field' galaxies as having $d_{\rm host} >$ 1.5\,Mpc, beyond which
$f_{\rm quenched}$ is constant.}
\end{figure}

We first explore the fraction of quenched dwarf galaxies as a function
of our environment parameter, $d_{\rm host}$. Figure~\ref{fig_environ}
shows the quenched fraction $f_{\rm quenched}$ for various bins in
stellar mass.  Quenched galaxies are preferentially found near a
massive host.  The observed quenched fraction decreases with
increasing values of $d_{\rm host}$, and then flattens to a constant
level for $d_{\rm host}$ distances greater than 1.5\,Mpc.  In our most
massive stellar bin, between $10^{9.5} -10^{10}$\Msun, the quenched
fraction levels off at a value of $f_{\rm quenched} = 5$\% at large
distances from a host galaxy.  For stellar mass less than $\sim
10^{9}$\Msun, the quenched fractions beyond 1.5\,Mpc are effectively
zero (see \S\,\ref{ssec_noreds} for further discussion).  In dense
environments, $d_{\rm host} < 0.25$\,Mpc, the quenched fraction does
not appear to be a strong function of stellar mass, approaching $f_{\rm
  quenched} = 30$\% in all our stellar mass bins.

Based on Figure~\ref{fig_environ}, we define ``field'' or ``isolated''
galaxies as those with $d_{\rm host} > 1.5$\,Mpc.  Our definition is
motivated by Figure~\ref{fig_environ}: we see evidence of
environmental processes, in the form of increased quenched fractions,
out to at least 1\,Mpc.  The host galaxies in our sample cover a range
of masses: as calculated in \S\,\ref{ssec_virial}, the virial radii of
our host galaxies range between $150-1000$\,kpc, with a median virial
radius of 225\,kpc.  Thus, we are seeing evidence for environment processes
several times beyond the host's virial radii.  This result is consistent with
\citet{wang09a} and we discuss implications of
this statement in \S\,\ref{ssec_virial}.

In the sections below, we focus on the properties of field galaxies.
Our measured quenched fractions for field dwarf galaxies with stellar
mass less than $10^9$\Msun\ are effectively zero, while published red
fractions in the same stellar mass range, based only on $g-r$ colors,
are between 5-10\% of isolated galaxies \citep{wang09a,peng11a}.
Using colors only, we reproduce the Wang et al.~result (Wang et
al.~Figure~2), finding a red (as opposed to quenched) fraction of 6\%
in the mass range $10^8 - 10^9$\Msun\ beyond 1.5\,Mpc.  We find that
the majority of these red galaxies (46 out of 49) have strong
H$\alpha$ emission; none of these galaxies would not pass our combined
H$\alpha$ plus D$_n$4000 criteria for quenching.  The distribution of
axis ratios (b/a, based on the 2D Sersic $r$-band profile fits) for
blue galaxies in this stellar mass range is peaked at 0.5, as expected
for a population of disky systems viewed at random viewing positions.
The 49 isolated red star forming systems are preferentially disky,
peaking at an axis-ratio of 0.32, implying that most of these objects
are edge-on star-forming galaxies which appear red due to
dust-reddening.  This confirms that color selection alone is not a
good indicator in selecting quenched galaxies and motivates our
spectroscopic criteria.

\subsection{A Stellar Mass Threshold for Quenched Field Galaxies}
\label{ssec_thresh}

Previous studies have shown that massive galaxies which are the
central galaxy in their dark matter halo are predominantly quenched:
the quenched fractions of central galaxies is 100\% for galaxies with
stellar mass above $10^{11}$\Msun, decreasing to 20\% of galaxies with
stellar mass of $10^{10}$\Msun \citep{wetzel10a,peng11a, woo12a}.
None of these SDSS-based studies include galaxies less massive than
$10^{9}$\Msun.  With our cleaned SDSS dwarf galaxy sample, we next ask
whether the quenched fractions of central galaxies continue to
decrease with stellar mass and if there is a threshold below which
this fraction reaches zero.

We compare our dwarf galaxy sample to a modified version of the
group catalog from \citet{tinker11a} based on the SDSS DR7.  This is a
volume-limited ($z\le 0.06$) catalog including galaxies brighter than
$M_r = -18$.  To ensure homogeneity with our dwarf galaxy sample, we
remeasure physical properties of galaxies in the Tinker et
al.~catalog, using spectral measurements from \citet{yan11a} and
stellar masses from the NYU-VAGC.  We define quenched galaxies using
the same criteria as our dwarf sample: EW H$\alpha < 2$\mbox{\AA} and
D$_n$4000$ > 0.6 + 0.1$log$_{10}$[M$_{\rm stellar}$].

\begin{figure}[t!]
\epsscale{1.2}
\plotone{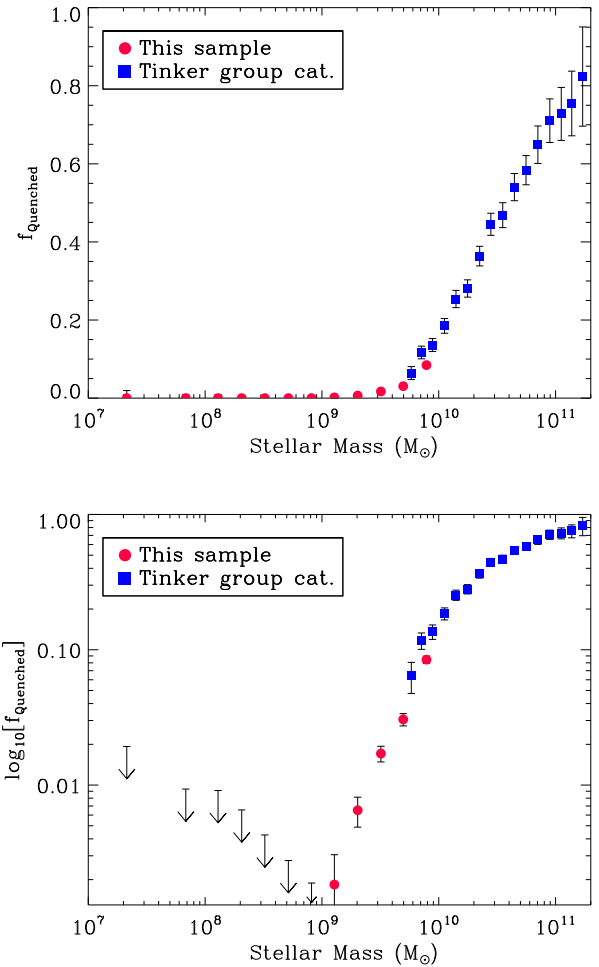}
\caption{\label{fig_f_red} The quenched fraction of isolated or
  central galaxies as a function of stellar mass.  Our sample of dwarf
  galaxies (red circles) is compared to the group catalog of Tinker et
  al.~at higher masses (blue squares).  The quenched fraction is zero
  for stellar masses below $1.0\times 10^9$\Msun.  The top panel shows
  the fraction on a linear scale, while the bottom panel are the same
  data points shown on a logarithmic scale.   One-sigma upper limits
  are shown for bins in which no quenched central galaxies are detected.}
\end{figure}

The Tinker et al.~catalog includes galaxies with stellar mass down to
$10^{9.6}$\Msun, providing direct overlap with our NSA dwarf galaxy
sample.  We compare our isolated field dwarf galaxies to their sample
of central galaxies.  By definition, our isolated dwarf galaxies will
be central galaxies, but central galaxies are not necessarily
isolated.  \citet{tinker11a} define central galaxies as those that do
not exist within the halo radius of a larger halo.  The halo radius is
defined as the radius within which the average density is 200 times
the background density. We note this is larger than the virial radius
in $\Lambda$CDM, within which the average density corresponds to
$\sim$360 times the background density \citep[Eqn.~6;][]{bryan97a}.
The halo radius of a smaller galaxy may overlap with a larger galaxy,
but will not be considered a satellite until the smaller galaxy itself
is within the larger's halo radius.  This motivates us to use a more
restrictive definition of central galaxy.  We have shown in
Figure~\ref{fig_environ} that galaxies below a stellar mass of
$10^{10}$\Msun\ show environmental effects out to as much as 1\,Mpc
away from a massive galaxy, several times larger than the host galaxy
halo radius.  We therefore re-calculate the number of central galaxies
in the Tinker et al.~catalog, searching for associated objects within
three times the halo radius.  Using this definition, 75\% of central
galaxies in the Tinker et al.~catalog have $d_{\rm host} > 1.5$\,Mpc
in the overlapping stellar mass range of our dwarf galaxy catalog.
This means that our definition of field galaxies is slightly more
isolated than that of central galaxies in the Tinker et al.~group
catalog.  This can be seen as slightly lower quenched fractions in the
region of overlap shown in Figure~\ref{fig_f_red}.

In Figure~\ref{fig_f_red}, we plot the fraction of quenched galaxies
as a function of stellar mass for central/field galaxies.  At high
stellar masses, quenched galaxies make up the majority of the central
galaxy population.  While Tinker et al.~and others have shown that the
quenched fraction of central galaxies is 100\% for stellar masses
greater than $10^{11}$\Msun, our H$\alpha$ cut removes objects with
strong AGN activity, decreasing $f_{\rm quenched}$ at these masses.
The quenched fraction decreases with stellar mass, reaching zero at a
stellar mass between $1-2\times 10^9$\Msun.  The least massive
quenched field galaxy in our sample has a stellar mass
$1.02\times10^9$\Msun.  We therefore conclude that there is a
threshold of $1.0\times10^9$\Msun\ below which quenched galaxies
are not found in the field.  This threshold represents a fundamental
stellar mass scale.  Dwarf galaxies with stellar mass below this scale
cannot quench star formation on their own.  The threshold stellar mass
does not change significantly for reasonable variations of our
quenched definition.  We list in Table~1, the number of galaxies,
$1/V_{\rm max}$ corrections and quenched fractions for our NSA sample.

A stellar mass threshold of $1.0\times 10^9$\Msun, below which
isolated quenched galaxies do not exist, is consistent with
extrapolations of previously published work \citep{peng10a, wetzel10a}
and confirm, with a larger sample, the conclusions of
\citet{haines08a,haines07a} who found no isolated quenched galaxies in the SDSS
DR4 in the absolute magnitude range $-16 < M_r < -18$.  These authors
use somewhat different definitions of both 'isolated' and 'quenched',
but also conclude that there is an absence of low mass isolated
quenched galaxies.

\subsection{The Absence of Quenched Dwarf Galaxies in the
Field}\label{ssec_noreds}

We establish above that quenched central galaxies with stellar masses
below $1.0\times 10^9$\Msun\ do not exist in the field.  Dwarf
galaxies with stellar mass below this scale cannot quench star
formation on their own; {\it all} field galaxies in our sample below
this threshold are forming stars (see \S\,\ref{ssec_SF}).  We detect
\nfield field galaxies in the stellar mass range $10^7
-10^{9.0}$\Msun.  Accounting for 1/V$_{\rm max}$ corrections, we
calculate an upper 1-sigma limit on the quenched fraction of $f_{\rm
  quenched} < 0.0006$, or 0.06\% using \citet{gehrels86a}.  In denser
regions, defined here as $d_{\rm host} < 0.25$\,Mpc, we find that 148
out of 1504 galaxies are quenched in the same stellar mass regime,
corresponding to a quenched fraction of 23\% after volume corrections.
Thus, while quenched central galaxies with stellar mass less than $1.0\times
10^9$\Msun\ exist in dense environments, we do not find these objects
in the field.

\subsection{Quenched Dwarf Galaxies Within 4 $r_{virial}$ of a Massive
Host}
\label{ssec_virial}

We next investigate the distribution of quenched dwarf galaxies
relative to their host galaxy.  Our sample contains 223 quenched
galaxies below a stellar mass of $10^9$\Msun, all of which are within
1.5\,Mpc and 1000\kms\ of a more luminous host galaxy.  We calculate the
virial masses of the hosts using the prescriptions of
\citet{behroozi10a} based on stellar mass.  We calculate a virial mass
for each host galaxy and determine its virial radius ($r_{\rm vir}$)
assuming the average enclosed density is 360 times the background
density.

\begin{figure}[t!]
\epsscale{1.2}
\plotone{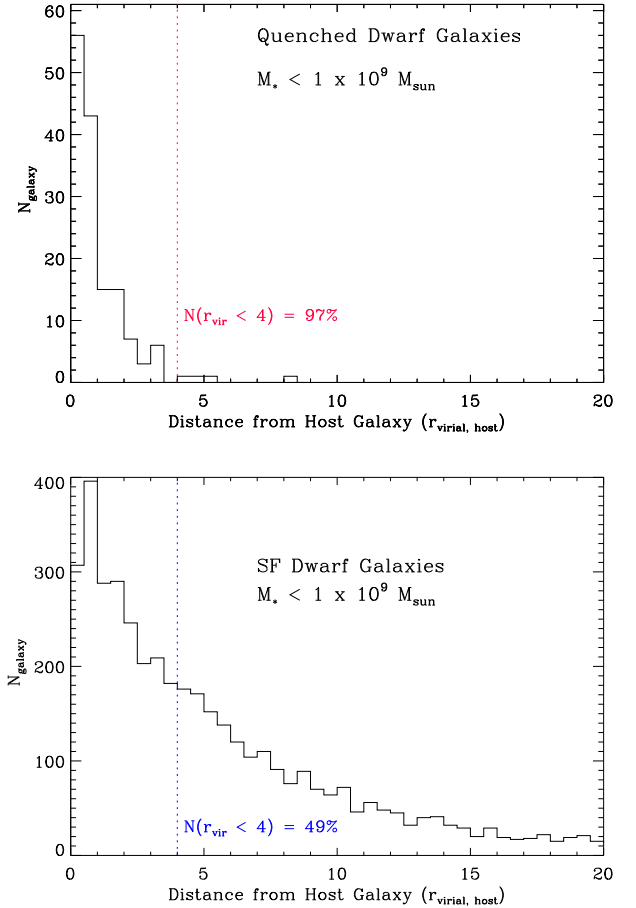}
\caption{\label{fig_rvir} The distribution of host distance for all
  quenched ({\it top}) and star forming ({\it bottom}) galaxies with
  stellar mass below $1\times10^9$\Msun.  Distances are plotted in
  units of host galaxy virial radius.  97\% of quenched dwarf galaxies
  are found within 4 virial radii (red dotted line), while only 49\%
  of star forming galaxies are found within a similar distance (blue
  dotted line).  The most distant quenched dwarf galaxy in our sample
  is 8\,$r_{\rm vir}$ from its host, while the most distant star
  forming objects (not shown in this figure) are over 50\,$r_{\rm
    vir}$ from a massive host galaxy.}
\end{figure}

In the top panel of Figure~\ref{fig_rvir}, we show the distribution of
quenched galaxies with stellar mass below $1\times10^9$\Msun\ as a
function of distance from its host galaxy, in units of the host's
virial radius.  In the bottom panel of Figure~\ref{fig_rvir}, we
compare this distribution to that of star forming dwarf galaxies
(defined as galaxies with detected H$\alpha > 2$\mbox{\AA}, see
\S\,\ref{ssec_SF}) in the same stellar mass range.  The majority
(87\%) of quenched dwarf galaxies are within 2\,$r_{\rm vir}$ of a
massive host galaxy and would thus be considered ``satellite''
galaxies, while 97\% of objects are within 4\,$r_{\rm vir}$.  For
comparison, less than 50\% of star forming dwarf galaxies are within
4\,$r_{\rm vir}$ of a massive neighbor.  The furthest quenched galaxy
is 8 $r_{\rm vir}$ from its host, while the furthest star forming
dwarf galaxy is over 50\,$r_{\rm vir}$ from a massive neighbor.

There are numerous proposed mechanisms to quench satellite galaxies
within the virial radius of a massive galaxy.  Processes such as ram
pressure stripping or tidally induced star formation can quickly
remove or use up gas as a satellite enters the virial radius of a
massive host.  The handful of quenched dwarf galaxies between
2-8\,$r_{\rm vir}$ may be evidence for quenching processes which act
at larger distances from the primary halo.  Alternatively, these may
be ``backsplash'' galaxies which have previously been within the host
virial radius, but are on either highly eccentric orbits or have been
dynamical ejected from the host halo \citep{ludlow09a, wang09a}.
Numerical simulations suggest that up to 10\% of satellites associated
with a massive galaxy host can reside as far as four virial radii
away, consistent with Figure~\ref{fig_rvir}.  These galaxies are
analogous to the dwarf spheroidal galaxies Cetus and Tucana (M$_{\rm
  stellar}\sim10^6$\Msun), which are roughly 1\,Mpc, or 3$-4r_{\rm
  vir}$ from the Milky Way \citep{fraternali09a}.

Our host galaxy definition is insensitive to whether or not a dwarf
galaxy has a companion with stellar mass less than $10^{10}$\Msun.
While we do not include dwarf galaxies with obvious signs of active
merging activity, there are a number of dwarf-dwarf galaxy pairs in
the SDSS DR8 volume.  We perform a cursory search for isolated dwarf
pairs, looking for two or more galaxies with stellar mass less than
$10^9$\Msun\ within a projected distance of 100\,kpc, a velocity
difference less than 100\kms, and are further than 1.5\,Mpc from a
massive host galaxy.  We find 39 pairs matching these criteria.
Significant H$\alpha$ is detected in all these dwarf galaxy pairs,
with a median H$\alpha$ EW of 40\mbox{\AA} (slightly above the median
of the full star-forming sample).  None of the galaxies in these dwarf
pairs are quenched.  The fact that quenched dwarf galaxies are only
found within 8\,$r_{\rm vir}$ of a more massive neighbor galaxy
suggests that such dwarf-only groups are not effective at quenching
their member dwarf galaxies.  We will explore the properties of
dwarf-dwarf systems in a future contribution.

\subsection{Continuous Star Formation in Isolated Dwarf Galaxies}
\label{ssec_SF}

We identify \nfield galaxies with stellar mass less than $1\times
10^9$\Msun\ that are in the field (as defined here, these are also
central/isolated galaxies).  All of our field dwarf galaxies show
evidence for recent star formation.  We briefly examine the
properties of these galaxies, deferring a full analysis for a future
paper.

The vast majority of field dwarf galaxies have detected H$\alpha$
gflux: 2940 out of \nfield (99.6\%) with stellar mass less than
$10^9$\Msun\ have H$\alpha$ EW $> 2$\mbox{\AA}, with a median EW of
32\,\mbox{\AA}.  The presence of H$\alpha$ emission implies star
formation within the past 50\,Myr \citep{bruzual03a}.  The median star
formation rate for these isolated galaxies is 0.03\,\Msun yr$^{-1}$,
based on FUV GALEX fluxes \citep{salim07a}.  This rate is comparable
to that expected by extrapolating the star
formation rate versus stellar mass relationship seen at higher stellar
masses \citep[e.g.,][]{wuyts11a}.  We note that none of these galaxies
show evidence for AGN activity based on their position in
[OIII]/H$\beta$ versus [NII]/H$\alpha$ line ratio diagnostic plot
\citep{kewley01a, yan11a}.  The overwhelming fraction of isolated
galaxies with very recent star formation suggests that this population
is in a state of continuous star formation.

There are 11 field dwarf galaxies (0.4\% of the population) which do
not show H$\alpha$ emission.  These objects have D$_n4000=1.1-1.3$,
suggesting a luminosity-weighted stellar age between 100-200\,Myr
based on \citet{bruzual03a} models.  We have examined the broad band
SDSS images and spectra of these 11 galaxies: 9 are strong
post-starburst galaxies (e.g., K+A galaxies) according to the criteria
defined by \citet{yan09a} through spectral decomposition, and two are
border line cases, satisfying the more inclusive K+A criteria defined
by \citet{balogh99a} based on H$\delta$ absorption EW.  We conclude
that these galaxies have shut off star formation within the past few
hundred million years, and less than 1 Gyr even for the two weaker
K+As. The fact that we do not find old quenched galaxies in the field
suggest that these rare occasions of post-starburst galaxies are a
transient phase in the life of isolated dwarf galaxies: they will soon
have star formation again.  This population constrains the
``burstiness'' of star formation, suggesting that star formation
rarely shuts off in isolated dwarf galaxies and, if so, for less than
a few hundred million years.

\section{Discussion and Conclusions}\label{sec_concl}

We demonstrate that quenched dwarf galaxies are rare in
isolation, existing almost exclusively in the vicinity of a more
massive neighbor.  For field galaxies, we find a stellar mass
threshold of $1.0\times 10^9$\Msun\ below which quenched galaxies do not
exist.   Most quenched galaxies below this stellar mass threshold are
found within 2 $r_{\rm vir}$ of a massive host galaxy, and 97\% of all
quenched dwarf galaxies are found within 4 $r_{\rm vir}$.

With the SDSS DR8 data, we can state that quenched galaxies in the
field do not exist below $1.0 \times10^9$\,\Msun, however, we cannot test
this statement below $10^7$\,\Msun.  The SDSS does not contain
a sufficient number of galaxies below this stellar mass.  Studies of
dwarf galaxies within 10\,Mpc with stellar masses below $10^7$\,\Msun\
are consistent with our results:  low mass quenched galaxies exist only
within a few virial radius of the Milky Way, M31 or the nearby M81
group \citep{ grcevich09a,weisz11a}.  Below the detection limits of
any current survey, the quenched fractions of field galaxies may rise
again at extremely low stellar masses.  Ultra-faint galaxies ($M_V >
-5$, $M_{\rm stellar} < 10^4$\,\Msun) are currently detectable out to
only 50\,kpc in the Milky Way halo \citep{walsh09a}.  These galaxies
are thought to form their stars only before reionization \citep{brown12a}.  If
counterparts of the ultra-faint galaxies exist in the field, these
will appear as quenched objects.  Thus, the quenched fractions of
field galaxies may rise again at extremely low masses.

There are claims in the literature of isolated quenched dwarf
galaxies, however, there are no definitive examples which pass our
definition of isolated/field galaxies.  At slightly lower stellar mass
than the present sample, the dwarf spheroidal galaxies Cetus and
Tucana ($\sim10^6$\Msun), discussed in \S\,\ref{ssec_virial}, are
often cited as isolated quenched galaxies, but lie within 1.5\,Mpc (3
to 4 virial radius) of the Milky Way.  \citet{karachentseva11a} list
ten candidate isolated dwarf spheroidal galaxies which they identify
via visual searches.  Several of these candidates lie within the SDSS
footprint, but are well below the spectroscopic magnitude limits.
These objects were not detected in HI surveys, and have no measured
radial velocities or secure distance estimates.  Spectroscopic
follow-up of these candidates will be an excellent test of our
conclusions.  Finally, we note no contradiction with the results of
\citet{wang09a} who find a percentage of isolated dwarf galaxies with
red broad-band colors at similar stellar masses: we demonstrate in
\S\,\ref{ssec_environ} that these are dust-reddened star-forming
galaxies.

The absence of isolated quenched galaxies below $10^{9}$\,\Msun\
provides strong constraint on the internal feedback processes
regulating star formation.  We find that {\it all} galaxies below this
stellar mass threshold are forming stars in the field: 99.6\% of
isolated field galaxies have formed stars within the past 50\,Myr,
while the remaining galaxies have had star formation with the past few
million years.  A future paper will explore the properties of these
star forming field dwarf galaxies (Blanton et al.~in prep).  In
\citet{geha06b}, we presented HI follow-up for SDSS dwarf galaxies,
concluding that external processes were required to fully remove gas
from a dwarf galaxy.  Exploring the HI and other properties of
isolated dwarf galaxies above and below our stellar mass threshold
will provide insight into the internal physical mechanisms, such as
heating from supernova or active galactic nuclei, operating to quench
galaxies at higher masses.  Our stellar mass threshold provides a
strong boundary condition for any of these mechanisms to
completely shut off star formation in low mass galaxies.

\acknowledgments

MG acknowledges support from NSF grant AST-0908752 and the Alfred
P.~Sloan Foundation.  We acknowledge R.\,Dav\'e, J.\,Moustakas, F.\,van
den Bosch, R.\,Wechsler, A.\,Wetzel and B.\,Willman for productive
discussions.  Funding for the NASA-Sloan Atlas has been provided by
the NASA Astrophysics Data Analysis Program (08-ADP08-0072).  This
research has made use of the NASA/IPAC Extragalactic Database (NED)
which is operated by the Jet Propulsion Laboratory, California
Institute of Technology, under contract with the National Aeronautics
and Space Administration.

\begin{deluxetable}{lcccccc} \label{table_mask}
\tabletypesize{\scriptsize}
\tablecaption{Quenched Field Galaxy Fractions for NSA Dwarf Catalog}
\tablewidth{0pt}
\tablehead{
\colhead{$M_{\rm stellar}$} &
\colhead{N$_{\rm quenched}$} &
\colhead{N$_{\rm total}$} &
\colhead{$\Sigma$ (N$_{\rm quenched}/V_{\rm max}$)} &
\colhead{$\Sigma$ (N$_{\rm total}/V_{\rm max}$)} &
\colhead{f$_{\rm quenched}$} &
\colhead{$\sigma_{f_{\rm quenched}}$} \\
\colhead{[log$_{10}$(\Msun)]} &
\colhead{} &
\colhead{}&
\colhead{} &
\colhead{}&
\colhead{}&
\colhead{}
}
\startdata
   9.9 &  238 & 2843 & 0.000070 & 0.000826 &  0.084 &  0.0052\\
   9.7 &   71 & 2910 & 0.000027 & 0.000900 &  0.031 &  0.0032\\
   9.5 &   24 & 2620 & 0.000017 & 0.000991 &  0.017 &  0.0023\\
   9.3 &    5 & 2071 & 0.000007 & 0.001148 &  0.007 &  0.0016\\
   9.1 &    1 & 1594 & 0.000003 & 0.001397 &  0.002 &  0.0014\\
   8.9 &    0 &  999 & 0.000000 & 0.001435 &  0.000 &  0.0019\\
   8.7 &    0 &  686 & 0.000000 & 0.001653 &  0.000 &  0.0028\\
   8.5 &    0 &  448 & 0.000000 & 0.001803 &  0.000 &  0.0043\\
   8.3 &    0 &  296 & 0.000000 & 0.002142 &  0.000 &  0.0065\\
   8.1 &    0 &  215 & 0.000000 & 0.003076 &  0.000 &  0.0091\\
   7.8 &    0 &  210 & 0.000000 & 0.006973 &  0.000 &  0.0093\\
   7.3 &    0 &  105 & 0.000000 & 0.016505 &  0.000 &  0.0193
\enddata
\tablecomments{Quenched fractions for field galaxies with $d_{\rm
    host} >$ 1.5\,Mpc.  Bins are 0.2\,dex in stellar mass
  or larger to contain a minimum of 100 galaxies.}
\end{deluxetable}


\begin{thebibliography}{59}
\expandafter\ifx\csname natexlab\endcsname\relax\def\natexlab#1{#1}\fi

\bibitem[{{Aihara} {et~al.}(2011)}]{dr8}
{Aihara}, H. {et~al.} 2011, \apjs, 193, 29

\bibitem[{{Baldry} {et~al.}(2006){Baldry}, {Balogh}, {Bower}, {Glazebrook},
  {Nichol}, {Bamford}, \& {Budavari}}]{baldry06a}
{Baldry}, I.~K., {Balogh}, M.~L., {Bower}, R.~G., {Glazebrook}, K., {Nichol},
  R.~C., {Bamford}, S.~P., \& {Budavari}, T. 2006, \mnras, 373, 469

\bibitem[{{Balogh} {et~al.}(1999){Balogh}, {Morris}, {Yee}, {Carlberg}, \&
  {Ellingson}}]{balogh99a}
{Balogh}, M.~L., {Morris}, S.~L., {Yee}, H.~K.~C., {Carlberg}, R.~G., \&
  {Ellingson}, E. 1999, \apj, 527, 54

\bibitem[{{Behroozi} {et~al.}(2010){Behroozi}, {Conroy}, \&
  {Wechsler}}]{behroozi10a}
{Behroozi}, P.~S., {Conroy}, C., \& {Wechsler}, R.~H. 2010, \apj, 717, 379

\bibitem[{{Bell} {et~al.}(2004){Bell}, {Wolf}, {Meisenheimer}, {Rix}, {Borch},
  {Dye}, {Kleinheinrich}, {Wisotzki}, \& {McIntosh}}]{bell04a}
{Bell}, E.~F., {Wolf}, C., {Meisenheimer}, K., {Rix}, H.-W., {Borch}, A.,
  {Dye}, S., {Kleinheinrich}, M., {Wisotzki}, L., \& {McIntosh}, D.~H. 2004,
  \apj, 608, 752

\bibitem[{{Binggeli} {et~al.}(1985){Binggeli}, {Sandage}, \&
  {Tammann}}]{binggeli85b}
{Binggeli}, B., {Sandage}, A., \& {Tammann}, G.~A. 1985, \aj, 90, 1681

\bibitem[{{Blanton}(2006)}]{blanton06a}
{Blanton}, M.~R. 2006, \apj, 648, 268

\bibitem[{{Blanton} {et~al.}(2005{\natexlab{a}}){Blanton}, {Eisenstein},
  {Hogg}, {Schlegel}, \& {Brinkmann}}]{blanton05b}
{Blanton}, M.~R., {Eisenstein}, D., {Hogg}, D.~W., {Schlegel}, D.~J., \&
  {Brinkmann}, J. 2005{\natexlab{a}}, \apj, 629, 143

\bibitem[{{Blanton} {et~al.}(2011){Blanton}, {Kazin}, {Muna}, {Weaver}, \&
  {Price-Whelan}}]{blanton11a}
{Blanton}, M.~R., {Kazin}, E., {Muna}, D., {Weaver}, B.~A., \& {Price-Whelan},
  A. 2011, \aj, 142, 31

\bibitem[{{Blanton} \& {Moustakas}(2009)}]{blanton09a}
{Blanton}, M.~R. \& {Moustakas}, J. 2009, \araa, 47, 159

\bibitem[{{Blanton} \& {Roweis}(2007)}]{blanton06b}
{Blanton}, M.~R. \& {Roweis}, S. 2007, \aj, 133, 734

\bibitem[{{Blanton} {et~al.}(2005{\natexlab{b}})}]{blanton05a}
{Blanton}, M.~R. {et~al.} 2005{\natexlab{b}}, \aj, 129, 2562

\bibitem[{Blanton {et~al.}(2005)}]{blanton04b}
Blanton, M.~R. {et~al.} 2005, \apj, 631, 208

\bibitem[{{Brown} {et~al.}(2012){Brown}, {Tumlinson}, {Geha}, {Kirby},
  {VandenBerg}, {Mu{\~n}oz}, {Kalirai}, {Simon}, {Avila}, {Guhathakurta},
  {Renzini}, \& {Ferguson}}]{brown12a}
{Brown}, T.~M., {Tumlinson}, J., {Geha}, M., {Kirby}, E.~N., {VandenBerg},
  D.~A., {Mu{\~n}oz}, R.~R., {Kalirai}, J.~S., {Simon}, J.~D., {Avila}, R.~J.,
  {Guhathakurta}, P., {Renzini}, A., \& {Ferguson}, H.~C. 2012, \apjl, 753, L21

\bibitem[{{Bruzual} \& {Charlot}(2003)}]{bruzual03a}
{Bruzual}, G. \& {Charlot}, S. 2003, \mnras, 344, 1000

\bibitem[{{Bryan} \& {Norman}(1998)}]{bryan97a}
{Bryan}, G.~L. \& {Norman}, M.~L. 1998, \apj, 495, 80

\bibitem[{{Chabrier}(2003)}]{chabrier03a}
{Chabrier}, G. 2003, \pasp, 115, 763

\bibitem[{{Colless} {et~al.}(2001)}]{colless01a}
{Colless}, M. {et~al.} 2001, \mnras, 328, 1039

\bibitem[{{Cooper} {et~al.}(2007){Cooper}, {Newman}, {Coil}, {Croton}, {Gerke},
  {Yan}, {Davis}, {Faber}, {Guhathakurta}, {Koo}, {Weiner}, \&
  {Willmer}}]{cooper07a}
{Cooper}, M.~C., {Newman}, J.~A., {Coil}, A.~L., {Croton}, D.~J., {Gerke},
  B.~F., {Yan}, R., {Davis}, M., {Faber}, S.~M., {Guhathakurta}, P., {Koo},
  D.~C., {Weiner}, B.~J., \& {Willmer}, C.~N.~A. 2007, \mnras, 376, 1445

\bibitem[{{Croton} {et~al.}(2006)}]{croton06a}
{Croton}, D.~J. {et~al.} 2006, \mnras, 365, 11

\bibitem[{{Dekel} \& {Birnboim}(2008)}]{dekel08a}
{Dekel}, A. \& {Birnboim}, Y. 2008, \mnras, 383, 119

\bibitem[{{Einasto} {et~al.}(1974){Einasto}, {Saar}, {Kaasik}, \&
  {Chernin}}]{einasto74a}
{Einasto}, J., {Saar}, E., {Kaasik}, A., \& {Chernin}, A.~D. 1974, \nat, 252,
  111

\bibitem[{{Ferguson} \& {Sandage}(1991)}]{ferguson91a}
{Ferguson}, H.~C. \& {Sandage}, A. 1991, \aj, 101, 765

\bibitem[{{Fraternali} {et~al.}(2009){Fraternali}, {Tolstoy}, {Irwin}, \&
  {Cole}}]{fraternali09a}
{Fraternali}, F., {Tolstoy}, E., {Irwin}, M.~J., \& {Cole}, A.~A. 2009, \aap,
  499, 121

\bibitem[{{Gabor} \& {Dav{\'e}}(2012)}]{gabor12a}
{Gabor}, J.~M. \& {Dav{\'e}}, R. 2012, astro-ph/1202.5315

\bibitem[{{Geha} {et~al.}(2006){Geha}, {Blanton}, {Masjedi}, \&
  {West}}]{geha06b}
{Geha}, M., {Blanton}, M.~R., {Masjedi}, M., \& {West}, A.~A. 2006, \apj, 653,
  240

\bibitem[{{Gehrels}(1986)}]{gehrels86a}
{Gehrels}, N. 1986, \apj, 303, 336

\bibitem[{{Giovanelli} {et~al.}(2005)}]{giovanelli05a}
{Giovanelli}, R. {et~al.} 2005, \aj, 130, 2598

\bibitem[{{Grcevich} \& {Putman}(2009)}]{grcevich09a}
{Grcevich}, J. \& {Putman}, M.~E. 2009, \apj, 696, 385

\bibitem[{{Haines} {et~al.}(2007){Haines}, {Gargiulo}, {La Barbera},
  {Mercurio}, {Merluzzi}, \& {Busarello}}]{haines07a}
{Haines}, C.~P., {Gargiulo}, A., {La Barbera}, F., {Mercurio}, A., {Merluzzi},
  P., \& {Busarello}, G. 2007, \mnras, 381, 7

\bibitem[{{Haines} {et~al.}(2008){Haines}, {Gargiulo}, \&
  {Merluzzi}}]{haines08a}
{Haines}, C.~P., {Gargiulo}, A., \& {Merluzzi}, P. 2008, \mnras, 385, 1201

\bibitem[{{Jones} {et~al.}(2004)}]{jones04a}
{Jones}, D.~H. {et~al.} 2004, \mnras, 355, 747

\bibitem[{{Karachentseva} {et~al.}(2011){Karachentseva}, {Karachentsev}, \&
  {Sharina}}]{karachentseva11a}
{Karachentseva}, V.~E., {Karachentsev}, I.~D., \& {Sharina}, M.~E. 2011,
  astro-ph/1104.2506

\bibitem[{{Kauffmann} {et~al.}(2003){Kauffmann}, {Heckman}, {White}, {Charlot},
  {Tremonti}, {Peng}, {Seibert}, {Brinkmann}, {Nichol}, {SubbaRao}, \&
  {York}}]{kauffman03b}
{Kauffmann}, G., {Heckman}, T.~M., {White}, S.~D.~M., {Charlot}, S.,
  {Tremonti}, C., {Peng}, E.~W., {Seibert}, M., {Brinkmann}, J., {Nichol},
  R.~C., {SubbaRao}, M., \& {York}, D. 2003, \mnras, 341, 54

\bibitem[{{Kauffmann} {et~al.}(2004){Kauffmann}, {White}, {Heckman},
  {M{\'e}nard}, {Brinchmann}, {Charlot}, {Tremonti}, \&
  {Brinkmann}}]{kauffmann04a}
{Kauffmann}, G., {White}, S.~D.~M., {Heckman}, T.~M., {M{\'e}nard}, B.,
  {Brinchmann}, J., {Charlot}, S., {Tremonti}, C., \& {Brinkmann}, J. 2004,
  \mnras, 353, 713

\bibitem[{{Kere{\v s}} {et~al.}(2005){Kere{\v s}}, {Katz}, {Weinberg}, \&
  {Dav{\'e}}}]{keres05a}
{Kere{\v s}}, D., {Katz}, N., {Weinberg}, D.~H., \& {Dav{\'e}}, R. 2005,
  \mnras, 363, 2

\bibitem[{{Kewley} {et~al.}(2001){Kewley}, {Dopita}, {Sutherland}, {Heisler},
  \& {Trevena}}]{kewley01a}
{Kewley}, L.~J., {Dopita}, M.~A., {Sutherland}, R.~S., {Heisler}, C.~A., \&
  {Trevena}, J. 2001, \apj, 556, 121

\bibitem[{{Kimm} {et~al.}(2009){Kimm}, {Somerville}, {Yi}, {van den Bosch},
  {Salim}, {Fontanot}, {Monaco}, {Mo}, {Pasquali}, {Rich}, \& {Yang}}]{kimm09a}
{Kimm}, T., {Somerville}, R.~S., {Yi}, S.~K., {van den Bosch}, F.~C., {Salim},
  S., {Fontanot}, F., {Monaco}, P., {Mo}, H., {Pasquali}, A., {Rich}, R.~M., \&
  {Yang}, X. 2009, \mnras, 394, 1131

\bibitem[{{Lisker} {et~al.}(2007){Lisker}, {Grebel}, {Binggeli}, \&
  {Glatt}}]{lisker07a}
{Lisker}, T., {Grebel}, E.~K., {Binggeli}, B., \& {Glatt}, K. 2007, \apj, 660,
  1186

\bibitem[{{Ludlow} {et~al.}(2009){Ludlow}, {Navarro}, {Springel}, {Jenkins},
  {Frenk}, \& {Helmi}}]{ludlow09a}
{Ludlow}, A.~D., {Navarro}, J.~F., {Springel}, V., {Jenkins}, A., {Frenk},
  C.~S., \& {Helmi}, A. 2009, \apj, 692, 931

\bibitem[{{Mateo}(1998)}]{mateo98a}
{Mateo}, M.~L. 1998, \araa, 36, 435

\bibitem[{{Peng} {et~al.}(2011){Peng}, {Lilly}, {Renzini}, \&
  {Carollo}}]{peng11a}
{Peng}, Y., {Lilly}, S.~J., {Renzini}, A., \& {Carollo}, M. 2011,
  astro-ph/1106.2546

\bibitem[{{Peng} {et~al.}(2010)}]{peng10a}
{Peng}, Y. {et~al.} 2010, \apj, 721, 193

\bibitem[{{Salim} {et~al.}(2007)}]{salim07a}
{Salim}, S. {et~al.} 2007, \apjs, 173, 267

\bibitem[{{Tanaka} {et~al.}(2005){Tanaka}, {Kodama}, {Arimoto}, {Okamura},
  {Umetsu}, {Shimasaku}, {Tanaka}, \& {Yamada}}]{tanaka05a}
{Tanaka}, M., {Kodama}, T., {Arimoto}, N., {Okamura}, S., {Umetsu}, K.,
  {Shimasaku}, K., {Tanaka}, I., \& {Yamada}, T. 2005, \mnras, 362, 268

\bibitem[{{Tinker} {et~al.}(2011){Tinker}, {Wetzel}, \& {Conroy}}]{tinker11a}
{Tinker}, J., {Wetzel}, A., \& {Conroy}, C. 2011, astro-ph/1107.5046

\bibitem[{{van den Bosch} {et~al.}(2008){van den Bosch}, {Aquino}, {Yang},
  {Mo}, {Pasquali}, {McIntosh}, {Weinmann}, \& {Kang}}]{vandenbosch08a}
{van den Bosch}, F.~C., {Aquino}, D., {Yang}, X., {Mo}, H.~J., {Pasquali}, A.,
  {McIntosh}, D.~H., {Weinmann}, S.~M., \& {Kang}, X. 2008, \mnras, 387, 79

\bibitem[{{Walsh} {et~al.}(2009){Walsh}, {Willman}, \& {Jerjen}}]{walsh09a}
{Walsh}, S.~M., {Willman}, B., \& {Jerjen}, H. 2009, \aj, 137, 450

\bibitem[{{Wang} {et~al.}(2009){Wang}, {Yang}, {Mo}, {van den Bosch}, {Katz},
  {Pasquali}, {McIntosh}, \& {Weinmann}}]{wang09a}
{Wang}, Y., {Yang}, X., {Mo}, H.~J., {van den Bosch}, F.~C., {Katz}, N.,
  {Pasquali}, A., {McIntosh}, D.~H., \& {Weinmann}, S.~M. 2009, \apj, 697, 247

\bibitem[{{Weisz} {et~al.}(2011){Weisz}, {Dalcanton}, {Williams}, {Gilbert},
  {Skillman}, {Seth}, {Dolphin}, {McQuinn}, {Gogarten}, {Holtzman}, {Rosema},
  {Cole}, {Karachentsev}, \& {Zaritsky}}]{weisz11a}
{Weisz}, D.~R., {Dalcanton}, J.~J., {Williams}, B.~F., {Gilbert}, K.~M.,
  {Skillman}, E.~D., {Seth}, A.~C., {Dolphin}, A.~E., {McQuinn}, K.~B.~W.,
  {Gogarten}, S.~M., {Holtzman}, J., {Rosema}, K., {Cole}, A., {Karachentsev},
  I.~D., \& {Zaritsky}, D. 2011, \apj, 739, 5

\bibitem[{{Wetzel} {et~al.}(2011){Wetzel}, {Tinker}, \& {Conroy}}]{wetzel10a}
{Wetzel}, A.~R., {Tinker}, J.~L., \& {Conroy}, C. 2011, astro-ph/1107.5311

\bibitem[{{Willick} {et~al.}(1997){Willick}, {Strauss}, {Dekel}, \&
  {Kolatt}}]{willick97a}
{Willick}, J.~A., {Strauss}, M.~A., {Dekel}, A., \& {Kolatt}, T. 1997, \apj,
  486, 629

\bibitem[{{Woo} {et~al.}(2012){Woo}, {Dekel}, {Faber}, {Noeske}, {Koo},
  {Gerke}, {Cooper}, {Salim}, {Dutton}, {Newman}, {Weiner}, {Bundy}, {Willmer},
  {Davis}, \& {Yan}}]{woo12a}
{Woo}, J., {Dekel}, A., {Faber}, S.~M., {Noeske}, K., {Koo}, D.~C., {Gerke},
  B.~F., {Cooper}, M.~C., {Salim}, S., {Dutton}, A.~A., {Newman}, J., {Weiner},
  B.~J., {Bundy}, K., {Willmer}, C.~N.~A., {Davis}, M., \& {Yan}, R. 2012,
  astro-ph/1203.1625

\bibitem[{{Wuyts} {et~al.}(2011)}]{wuyts11a}
{Wuyts}, S. {et~al.} 2011, \apj, 742, 96

\bibitem[{{Yan}(2011)}]{yan11b}
{Yan}, R. 2011, \aj, 142, 153

\bibitem[{{Yan} \& {Blanton}(2012)}]{yan11a}
{Yan}, R. \& {Blanton}, M.~R. 2012, \apj, 747, 61

\bibitem[{{Yan} {et~al.}(2009){Yan}, {Newman}, {Faber}, {Coil}, {Cooper},
  {Davis}, {Weiner}, {Gerke}, \& {Koo}}]{yan09a}
{Yan}, R., {Newman}, J.~A., {Faber}, S.~M., {Coil}, A.~L., {Cooper}, M.~C.,
  {Davis}, M., {Weiner}, B.~J., {Gerke}, B.~F., \& {Koo}, D.~C. 2009, \mnras,
  398, 735

\bibitem[{{Yang} {et~al.}(2007){Yang}, {Mo}, {van den Bosch}, {Pasquali}, {Li},
  \& {Barden}}]{yang07a}
{Yang}, X., {Mo}, H.~J., {van den Bosch}, F.~C., {Pasquali}, A., {Li}, C., \&
  {Barden}, M. 2007, \apj, 671, 153

\bibitem[{{Zehavi} {et~al.}(2011){Zehavi}, {Zheng}, {Weinberg}, {Blanton},
  {Bahcall}, {Berlind}, {Brinkmann}, {Frieman}, {Gunn}, {Lupton}, {Nichol},
  {Percival}, {Schneider}, {Skibba}, {Strauss}, {Tegmark}, \&
  {York}}]{Zehavi11a}
{Zehavi}, I., {Zheng}, Z., {Weinberg}, D.~H., {Blanton}, M.~R., {Bahcall},
  N.~A., {Berlind}, A.~A., {Brinkmann}, J., {Frieman}, J.~A., {Gunn}, J.~E.,
  {Lupton}, R.~H., {Nichol}, R.~C., {Percival}, W.~J., {Schneider}, D.~P.,
  {Skibba}, R.~A., {Strauss}, M.~A., {Tegmark}, M., \& {York}, D.~G. 2011,
  \apj, 736, 59

\end{thebibliography}
\end{document}